\documentstyle[preprint,aps,epsf]{revtex}                  
\tightenlines
\begin{document}
\pagestyle{empty}                                      
\preprint{
\font\fortssbx=cmssbx10 scaled \magstep2
\hbox to \hsize{
\hfill$\raise .5cm\vtop{              
                \hbox{NCKU-HEP-98-14}\hbox{NCTU-HEP-9806}}$}
}
\draft
\vfill
\title{Perturbative QCD study of the $B\to K^*\gamma$ decay}
\vfill
\author{Hsiang-nan Li$^a$ and Guey-Lin Lin$^b$}

\address{$^a$\rm Department of Physics, National Cheng-Kung University,
Tainan 701, Taiwan, Republic of China}

\address{$^b$\rm Institute of Physics, National Chiao-Tung University,
Hsinchu 300, Taiwan, Republic of China}
%
%
\vfill
\maketitle
\begin{abstract}
We apply the perturbative QCD factorization theorem developed
recently for nonleptonic heavy meson decays 
to the radiative decay $B\to K^*\gamma$. In this
formalism the evolution of the Wilson coefficients from the $W$ boson mass
down to the characteristic scale of the decay process is governed by
the effective weak Hamiltonian. The evolution from the characteristic scale
to a lower hadronic scale is formulated by the Sudakov resummation. Besides 
computing
the dominant contribution arising from the magnetic-penguin operator $O_7$,
we also calculate the contributions of four-quark operators.  
By fitting our
prediction for the branching ratio of the $B\to K^*\gamma$ decay to the
CLEO data, we determine the $B$ meson wave function, that possesses a sharp
peak at a low momentum fraction.

\end{abstract}

\vskip 1.0cm

%
\pagestyle{plain}

\section{INTRODUCTION}

The observation of the decay $B\to K^*\gamma$ five years ago \cite{CLEOEX1} 
opened a new era for particle physics, since the penguin structure of the
electroweak theory was probed for the first time. Soon after this
observation, the inclusive $B\to X_s \gamma$ decay \cite{CLEOIC1} was also
established. The updated branching ratios for both decays are
$(4.2\pm 0.8\pm 0.6) \times 10^{-5}$ \cite{CLEOEX2} and 
$(3.14\pm 0.48)\times 10^{-4}$ \cite{CLEOIC2,ALEPH}, respectively. In recent
literatures, the inclusive decay $B\to X_s \gamma$ receives more
attentions than the exclusive mode $B\to K^*\gamma$. The branching ratio
and the photon energy spectrum of the $B\to X_s \gamma$ decay have been
used to constrain the parameter-space of the new physics beyond the standard
model. The preference of studying the inclusive mode is clear in that its
hadronic dynamics is much easier to handle as compared to the exclusive
mode. It has been shown that, to the leading order in $1/M_b$, $M_b$ being
the $b$ quark mass, the branching ratio ${\cal B}(B\to X_s\gamma)$ 
is given by the
branching ratio ${\cal B}(b\to s\gamma)$ of the 
corresponding quark-level process.
Furthermore, the sub-leading $1/M_b$-corrections can be parametrized
systematically using the heavy quark effective theory \cite{FLS}.  

The hadronic dynamics of the exclusive decay $B\to K^*\gamma$ is much more
complicated. Specifically, one has to deal with the soft dynamics involved
in the $B$ and $K^*$ mesons. Since the final states are light compared to
the decaying $B$ meson, one anticipates that perturbative QCD (PQCD) is
applicable to this process because of the large energy release. In fact,
the PQCD formalism based upon factorization theorems, which
incorporates the Sudakov resummation of soft-gluon effects, has been
developed for sometime \cite{LY1,L1,LY3}, and applied to semileptonic
\cite{LY1,L1}, inclusive-radiative \cite{LY3} and nonleptonic $B$ meson
decays \cite{CL}. This approach is so far rather successful. In this article,
we shall extend this formalism to penguin-induced exclusive processes, such
as the $B\to K^*\gamma$ decay. The satisfactory result with repect to 
this complicated
process, as presented later, provides further confidence on the validity of
the PQCD approach to $B$ meson decays.

According to the PQCD factorization theorem, the branching ratio of a heavy
meson decay is expressed as the convolution of a hard subamplitude with
meson wave functions. The former, with at least one hard gluon
attaching to the spectator quark, is calculable in the
usual perturbation theory,
while the latter must be extracted from the experimental data or derived by
nonperturbative methods, such as QCD sum rules. Since the $K^*$ meson
wave function has been known from the sum rule analyses \cite{CZ,BC}, the
$B\to K^*\gamma$ decay is an ideal process from which the unknown $B$
meson wave function can be determined. With the $B$ meson wave function
obtained here, we are able to make predictions for other $B$ meson decay
modes, especially for the charmless decays. We shall show that by fitting our
prediction for the branching ratio ${\cal B}
(B\to K^*\gamma)$ to the CLEO data, a
$B$ meson wave function with a sharp peak at the low momentum fraction is
obtained.

Comparing our work to others, we remark that most of the
previous studies on this decay focus only on the contribution
of the magnetic-penguin operator $O_7$, and their approaches are based upon
quark models.   
To ensure that other contributions are indeed negligible, we also
calculate the contribution by the current-current operator $O_2$, which arises
through the four-point $b\to sg^*\gamma$ coupling with the off-shell gluon
reabsorbed by a spectator quark. Although such a contribution has been calculated  
before\cite{MIL}, it is however obtained in a naive PQCD framework which does not 
include 
the Sudakov resummation and the renormalization-group (RG) analysis
\cite{MIL}.  As a result, the predictions of Ref. \cite{MIL} 
are sensitive to the choice of the
renormalization scale $\mu$. In our more completed approach, the sensitivity to $\mu$
can be avoided.         
We find that, due to certain cancellations, the contribution by $O_2$ 
turns out to be 
rather small.   

We also like to comment on a different viewpoint based on the overlap integral
of meson wave functions \cite{GSW}, which showed that the diagrams without
any hard gluon dominate over those we will be considering here. We shall
argue that the observation in \cite{GSW} is due to an underestimation on  
the value of the strong coupling constant and a choice of a flat $B$ meson
wave function. If evaluating the coupling constant at the characteristic
momentum flow involved in the decay process \cite{MIL}, and employing a
sharper $B$ meson wave function, these higher-order contributions may
become comparable to the leading-order ones. Hence, the approach in
\cite{GSW} does not seem to be self-consistent. In our approach, the
diagrams without hard gluons do not contribute under our parametrization of
parton momenta. Therefore, the diagrams with a hard gluon attaching to a
spectator are leading in our analysis.

This article is organized as follows: In Sect. II, we write down the
effective Hamiltonian for the $B\to K^* \gamma$ decay. In particular, 
the current-current operator $O_2$ and the magnetic-penguin operators
$O_7$ are identified as major sources of the contribution.
We then calculate the $O_2$-induced $b\to sg^*\gamma$ vertex, 
keeping the gluon
line off-shell. In Sect. III, we derive the factorization formulas for
the $B\to K^*\gamma$ decay, which include contributions from the various
operators. The numerical result is presented in Sect. IV, where the
contribution of each operator is compared. Sect. V is the conclusion.
   
\section{EFFECTIVE HAMILTONIAN}

The effective Hamiltonian for the flavor-changing $b\to s$ transition is 
now standard, which is given by \cite{REVIEW,SIGN}
\begin{equation}
H_{\rm eff}(b\to s \gamma)=-{G_F\over
\sqrt{2}}V^*_{ts}V_{tb}\sum_{i=1}^{8}C_i(\mu)O_i(\mu)
\end{equation}
with
\begin{eqnarray}
O_1 &=& (\bar{s}_ic_j)_{V-A}(\bar{c}_jb_i)_{V-A} \nonumber \\
O_2 &=& (\bar{s}_ic_i)_{V-A}(\bar{c}_jb_j)_{V-A}  \nonumber \\
O_3 &=& (\bar{s}_ib_i)_{V-A}\sum_{q}(\bar{q}_jq_j)_{V-A}  \nonumber
\\
O_4 &=& (\bar{s}_ib_j)_{V-A}\sum_{q}(\bar{q}_jq_i)_{V-A}  \nonumber
\\
O_5 &=& (\bar{s}_ib_i)_{V-A}\sum_{q}(\bar{q}_jq_j)_{V+A}  \nonumber
\\
O_6 &=& (\bar{s}_ib_j)_{V-A}\sum_{q}(\bar{q}_jq_i)_{V+A}  \nonumber
\\
O_7 &=& {e\over
4\pi^2}\bar{s}_i\sigma^{\mu\nu}(m_sP_L+m_bP_R)b_iF_{\mu\nu} \nonumber
\\
O_8 &=& {g\over
4\pi^2}\bar{s}_i\sigma^{\mu\nu}(m_sP_L+m_bP_R)T^a_{ij}
b_jG^a_{\mu\nu},
\end{eqnarray}  
$i, \ j$ being the color indices. In the PQCD picture, the lowest-order
diagrams for the $B\to K^* \gamma$ decay arising from operators $O_2$,
$O_7$ and $O_8$ are depicted in Figs.~1-3. It is not hard to see that
contributions other than the above are negligible. For example,
contributions depicted in Fig.~4 are very suppressed, although they are of
the same order, $O(eG_F\alpha_s)$ \cite{ORDER}, as those of Figs.~1-3. 
We draw this conclusion from a previous experience with $B\to D^*\gamma$.
Indeed, from a diagram similar to Fig.~4 (with $s$ replaced by $c$, and
the $\bar{q}$ on both sides replaced by $\bar{d}$ and $\bar{u}$, 
respectively), 
one has obtained ${\cal B}(B\to D^*\gamma)=10^{-6}$ \cite{CHENG}. Since
$f_{K^*}\approx f_{D^*}$, and $V_{tb}V_{ts}$ is comparable to
$V_{cb}V_{ud}$, it is clear that the diagram in Fig. 4 gives
${\cal B}(B\to K^*\gamma)\approx 10^{-6}\times C_{3(4,5,6)}^2\approx
10^{-9}-10^{-10}$, and is thus negligible. There is still one more type of
contributions of the same order as shown in Fig. 5. By an explicit
calculation, one can show that such contributions merely give corrections
to the Wilson coefficients $C_3$-$C_6$ occuring in Fig.~4. 
Since the matrix elements of Fig.~4 and Fig.~5 have identical tensor 
structures,
the branching ratio contributed by the latter figure  
behaves like $10^{-6}\times C_2^2
\times ({\alpha_s\over \pi})^2\approx 10^{-9}$, which is
also negligible. Finally, in Sect. IV, 
we shall see that the contribution by 
$O_8$ is negligible as well.        

Before implementing the PQCD formalism to evaluate the contributions of
Figs.~1-3, we compute the four-point $b\to sg^*\gamma$ vertex. 
Our calculation essentially generalizes the work by Liu and Yao 
\cite{LY} to the off-shell gluon case\cite{COM}. 
We first
perform a Fierz transformation on $O_2$, {\it i.e.},
\begin{equation}
O_2\equiv(\bar{s}_i c_i)_{V-A}(\bar{c}_jb_j)_{V-A}=
(\bar{s}_i b_j)_{V-A}(\bar{c}_j c_i)_{V-A}.
\end{equation}
The $b(p)\to s(p')\gamma(k_1) g^*(k_2)$ vertex is expressed as 
\begin{equation}
I^{a\nu}=C_2V_{ts}^*V_{tb}\bar{u}(p'){1\over 2} \gamma^{\rho}
(1-\gamma_5)T^a u(p)I_{\mu\nu\rho}
\epsilon^{\mu}(k_1),
\end{equation}
with the structure-tensor
\begin{eqnarray}
I_{\mu\nu\rho}&=&A_1\epsilon_{\mu\nu\rho\sigma}k_1^{\sigma}
+A_2 \epsilon_{\mu\nu\rho\sigma}k_2^{\sigma}
+A_3\epsilon_{\mu\rho\sigma\tau}k_1^{\sigma}k_2^{\tau}k_{1\nu}\nonumber \\
&+& A_4\epsilon_{\nu\rho\sigma\tau}k_1^{\sigma}k_2^{\tau}k_{2\mu}
+A_5\epsilon_{\mu\rho\sigma\tau}k_1^{\sigma}k_2^{\tau}k_{2\nu}
+A_6\epsilon_{\nu\rho\sigma\tau}k_1^{\sigma}k_2^{\tau}k_{1\mu}.
\end{eqnarray}
Clearly, the form factor $A_6$ can be discarded because of $k_{1}
\cdot \epsilon(k_1)=0$. From the requirement of gauge invariance,
{\it i.e.}, $k_1^{\mu}I_{\mu\nu\rho}=0$ and $k_2^{\nu}I_{\mu\nu\rho}=0$,
we have 
\begin{eqnarray}
& & A_2+A_4 k_1\cdot k_2=0\nonumber \\
& & A_1+A_3 k_1\cdot k_2 +A_5 k_2^2=0.
\end{eqnarray}
The invariance of $I_{\mu\nu\rho}$ under the interchanges
$k_1\leftrightarrow k_2$ and $\mu \leftrightarrow \nu$ further require
$A_3=-A_4$. With the above relations, $I_{\mu\nu\rho}$ is simplified into
\begin{eqnarray}
I_{\mu\nu\rho}&=&A_4\left[k_1\cdot k_2\epsilon_{\mu\nu\rho\sigma}
(k_1-k_2)^{\sigma}+\epsilon_{\nu\rho\sigma\tau}k_1^{\sigma}k_2^{\tau}
k_{2\mu}-\epsilon_{\mu\rho\sigma\tau}k_1^{\sigma}k_2^{\tau}k_{1\nu}\right]
\nonumber \\
&+& A_5\left(\epsilon_{\mu\rho\sigma\tau}k_1^{\sigma}k_2^{\tau}
k_{2\nu}-k_2^2\epsilon_{\mu\nu\rho\sigma}k_1^{\sigma}\right),
\end{eqnarray}
with
\begin{eqnarray}
A_4&=&{2\sqrt{2}\over 3\pi^2}eg_sG_FI_{11}(M_c^2)\;,
\\
A_5&=&-{2\sqrt{2}\over 3\pi^2}eg_sG_F\left[I_{10}(M_c^2)-I_{20}(M_c^2)
\right]\;,
\end{eqnarray} 
and
\begin{equation}
I_{ab}(m^2)=\int_{0}^{1}{dx}\int_{0}^{1-x}{dy}{x^ay^b\over
x(1-x)k_2^2+2xyk_1\cdot k_2-m^2+i\varepsilon}.
\label{iab}
\end{equation}
Carrying out the Feynman-parameter integrations, we obtain
\begin{eqnarray}
I_{11}(m^2)&=& \left[{1\over 2Q^2}+\int_{0}^{1}{dx}{m^2-x(1-x)k_2^2
\over xQ^4}\ln \left| {m^2-x(1-x)(k_2^2+Q^2)\over m^2-x(1-x)k_2^2}\right|
 \right.
\nonumber \\
&-& \left. i\pi\Theta(Q^2+k_2^2-4m^2)\left({m^2\over Q^4}
\ln{1+\beta\over 1-\beta}-{\beta k_2^2\over 2Q^4}\right)\right]\;, 
\label{i11}\\
I_{10}(m^2)-I_{20}(m^2)&=&\left[\int_{0}^{1}{dx}{1-x\over Q^2}
\ln \left| {m^2-x(1-x)(k_2^2+Q^2)\over m^2-x(1-x)k_2^2}\right|\right.
\nonumber \\
&-& \left. i\pi\Theta(Q^2+k_2^2-4m^2){\beta\over 2Q^2}\right]\;,
\label{i10}
\end{eqnarray}  
with $Q^2=2k_1\cdot k_2$, and $\beta=\sqrt{1-4m^2/(k_2^2+Q^2)}$. 
As a side remark, we note that both $I_{11}(m^2)$ and 
$I_{10}(m^2)-I_{20}(m^2)$
have smooth limits as $m^2\to 0$. Hence, we can safely neglect 
the contributions of $O_4$ and $O_6$ to the $b\to sg^*\gamma$ vertex, since
their Wilson coefficients, $C_4$ and $C_6$, are much smaller than $C_2$.  

Having determined the $b\to sg^*\gamma$ vertex, we attach the off-shell
gluon line to the spectator quark to form the $B\to K^*\gamma$ amplitude
as shown in Fig.~1. As mentioned, this amplitude is of the order
$e\alpha_sG_F$, the same as the amplitudes induced by $O_7$ and $O_8$
depicted in Figs.~2 and 3, respectively. In the next section, we shall
compute the contribution of each operator to the $B\to K^*\gamma$ decay
using the PQCD factorization theorem. 

\section{FACTORIZATION FORMULAS}

In this section we first review the PQCD factorization theorem developed
for nonleptonic heavy meson decays \cite{CL}, and then extend it to the
$B\to K^*\gamma$ decay. Nonleptonic heavy meson decays involve three scales:
the $W$ boson mass $M_W$, at which the matching conditions of the effective
Hamiltonian to the original Hamiltonian are defined, the typical scale $t$
of a hard subamplitude, which reflects the dynamics of heavy meson decays,
and the factorization scale $1/b$, with $b$ the conjugate variable of parton
transverse momenta. The dynamics below $1/b$ is regarded as being
completely nonperturbative, and parametrized into a meson wave fucntion
$\phi(x)$, $x$ being the momentum fraction. Above the scale $1/b$, PQCD 
is reliable and radiative corrections produce two types of large logarithms:
$\ln(M_W/t)$ and $\ln(tb)$. The former are summed by RG equations to give
the evolution from $M_W$ down to $t$ described by the Wilson coefficients
$c(t)$. While the latter are summed to give the evolution from $t$ to $1/b$.

There exist also double logarithms $\ln^2(Pb)$ from the overlap of
collinear and soft divergences, $P$ being the dominant light-cone component
of a meson momentum. The resummation of these double logarithms leads to a
Sudakov form factor $\exp[-s(P,b)]$, which suppresses the long-distance
contributions in the large $b$ region, and vanishes as $b=1/\Lambda$,
$\Lambda\equiv \Lambda_{\rm QCD}$ being the QCD scale. This factor improves
the applicability of PQCD around the energy scale of few GeV. The $b$ quark
mass scale is located in the range of applicability. This is the motivation
we develop the PQCD formalism for heavy hadron decays. For the detailed
derivation of the relevant Sudakov form factors, please refer to
\cite{LY1,L1}.

With all the large logarithms organized, the remaining finite contributions
are absorbed into a hard $b$ quark decay subamplitude $H(t)$. Because of
Sudakov suppression, the perturbative expansion of $H$ in the coupling
constant $\alpha_s$ makes sense. Therefore, a three-scale factorization
formula is given by the typical expression, 
\begin{eqnarray}
c(t)\otimes H(t)\otimes \phi(x)
\otimes\exp\left[-s(P,b)-2\int_{1/b}^t\frac{d{\bar\mu}}
{\bar\mu}\gamma_q(\alpha_s({\bar\mu}))\right],
\label{for}
\end{eqnarray}
where the exponential containing the quark anomalous dimension 
$\gamma_q=-\alpha_s/\pi$ describes the evolution from $t$ to $1/b$ 
mentioned above. The explicit expression of the exponent $s$ is can be found
in \cite{CL}. Since logarithmic corrections have been summed by RG 
equations, the above factorization formula does
not depend on the renormalization scale $\mu$ \cite{CL}. Our formalism then
avoids the sensitivity to $\mu$ that appears in \cite{MIL}.

We now apply the three-scale factorization theorem to the radiative
decay $B\to K^*\gamma$, whose effective Hamiltonian has been given in the
previous section. As stated before, only the operators $O_2$, $O_7$ and
$O_8$ are crucial, to which the corresponding diagrams and hard
subamplitudes are shown in Figs.~1-3 and Table I, respectively.
We write the momenta of $B$ and $K^*$ mesons in light-cone coodinates as
$P_B=(M_B/\sqrt{2})(1,1,{\bf 0}_T)$ and
$P_K=(M_B/\sqrt{2})(1,r^2,{\bf 0}_T)$, respectively, with $r=M_{K^*}/M_B$.
The $B$ meson is at rest with the above choice of momenta. We further
parametrize the momenta of the light valence quarks in the $B$ and $K^*$
mesons as $k_B$ and $k_K$, respectively. $k_B$ has a minus component
$k_B^-$, giving the momentum fraction $x_B=k_B^-/P_B^-$, and small
transverse components ${\bf k}_{BT}$. $k_K$ has a large plus component
$k_K^+$, giving $x_K=k_K^+/P_K^+$, and small ${\bf k}_{KT}$. The photon
momentum is then $P_\gamma=P_B-P_K$, whose nonvanishing component is only
$P_\gamma^-$.

The $B\to K^*\gamma$ decay amplitude can be decomposed as
\begin{equation}
M=\epsilon^*_\gamma\cdot\epsilon^*_{K^*}M^S+
i\epsilon_{\mu\rho+-}\epsilon_\gamma^{*\mu}\epsilon^{*\rho}_{K^*}M^P\;,
\label{am}
\end{equation}
with $\epsilon_\gamma$ and $\epsilon_{K^*}$ the polarization vectors of the
photon and of the $K^*$ meson, respectively. 
Note that we have neglected the structure $(P_\gamma\cdot \epsilon^*_{K^*})
(P_K\cdot \epsilon^*_\gamma)/(P_\gamma\cdot P_K)$ which should come together 
with $\epsilon^*_\gamma\cdot\epsilon^*_{K^*}$. This is due to our choice
of the frame which gives $P_K\cdot \epsilon^*_\gamma=0$. From
Eq.~(\ref{am}), it is obvious that only the $K^*$ mesons with
transverse polarizations are produced in the decay.
 
The total rate of the $B\to K^*\gamma$ decay is given by
\begin{equation}
\Gamma=\frac{1-r^2}{8\pi M_B}(|M^S|^2+|M^P|^2)\;.
\end{equation}
We can further decompose $M^S$ and $M^P$ as
\begin{equation}
M^i=M^i_2+M^i_7+M^i_8 \;,
\end{equation}
where $i=S$ or $P$, and the terms on the right-hand side represent
contributions from operators $O_2$, $O_7$, and $O_8$, respectively. 

In the following, we write the factorization formulas for $M^i_l$
in terms of the overall factor
\begin{equation}
\Gamma^{(0)}=\frac{G_F}{\sqrt{2}}\frac{e}{\pi}V^*_{ts}V_{tb}C_FM_B^5\;.
\end{equation}
The amplitudes contributed by $O_2$ is written as
\begin{eqnarray}
M^S_2&=&\Gamma^{(0)}\frac{4}{3}\int_0^1dx\int_0^{1-x}dy\int_0^1dx_Bdx_K
\int_0^{1/\Lambda}bdb\phi_B(x_B)\phi_{K^*}(x_K)
\nonumber\\
& &\times \alpha_s(t_2)c_2(t_2)\exp[-S(x_B,x_K,t_2,b,b)]
\nonumber\\
& &\times[(1-r^2+2rx_K+2x_B)y-(rx_K+3x_B)(1-x)]
\nonumber\\
& &\times
\frac{(1-r)(1-r^2)x_Kx}{xy(1-r^2)x_KM_B^2-M_c^2}
H_{2}(Ab,\sqrt{|B_2^2|}b)\;,
\end{eqnarray}
\begin{eqnarray}
M^P_2&=&\Gamma^{(0)}\frac{4}{3}\int_0^1dx\int_0^{1-x}dy\int_0^1dx_Bdx_K
\int_0^{1/\Lambda}bdb\phi_B(x_B)\phi_{K^*}(x_K)
\nonumber\\
& &\times \alpha_s(t_2)c_2(t_2)\exp[-S(x_B,x_K,t_2,b,b)]
\nonumber\\
& &\times \left[\left((1-r)(1-r^2)+2r^2x_K+2x_B\right)y\right.
\nonumber\\
& &\left.-\left(r(1+r)x_K+(3-r)x_B\right)(1-x)\right]
\nonumber\\
& &\times 
\frac{(1-r^2)x_Kx}{xy(1-r^2)x_KM_B^2-M_c^2}H_{2}(Ab,\sqrt{|B_2^2|}b)\;,
\end{eqnarray}
with
\begin{eqnarray}
A^2&=&x_Kx_BM_B^2\;,
\nonumber\\
B_2^2&=&x_Kx_BM_B^2-\frac{y}{1-x}(1-r^2)x_KM_B^2+\frac{M_c^2}{x(1-x)}\;,
\nonumber\\
t_2&=&\max(A,\sqrt{|B_2^2|},1/b)\;.
\end{eqnarray}
To arrive at the above expressions, we have employed Eq.~(\ref{iab}) for
the charm loop integral, instead of Eqs.~(\ref{i11}) and (\ref{i10}). The
variable $b_B$ ($b_K$), conjugate to the parton transverse momentum
$k_{BT}$ ($k_{KT}$), represents the transverse extent of the $B$ ($K^*$)
meson; $t_2$ is the characteristic scale of the hard subamplitude
\begin{eqnarray}
H_{2}(Ab,\sqrt{|B_2^2|}b)&=&K_0(Ab)-K_0(\sqrt{|B_2^2|}b)\;,\;\;\;\;
B_2^2>0\;,
\nonumber\\
&=&K_0(Ab)-i\frac{\pi}{2}H_0^{(1)}(\sqrt{|B_2^2|}b)\;,\;\;\;\;
B_2^2<0\;,
\end{eqnarray}
which comes from the Fourier transform of the corresponding expressions
in Table I to the $b$ space. Note that, for simplicity in the notation,
we have used 
$H_2$ to denote hard subamplitudes in both $M_2^S$ and $M_2^P$. In Table I,
these two amplitudes are distinguished.

There are two diagrams, Figs.~2(a) and 2(b), associated with the operator
$O_7$, where the hard gluon connects both quarks of the $B$ meson
or those of the $K^*$ meson. The amplitudes from the two diagrams are
\begin{eqnarray}
M^S_7=-M^P_7&=&\Gamma^{(0)}2\int_0^1dx_Bdx_K\int_0^{1/\Lambda}
b_Bdb_Bb_Kdb_K\phi_B(x_B)\phi_{K^*}(x_K)(1-r^2)
\nonumber\\
& &\times\left[rH^{(a)}_{7}(Ab_K,\sqrt{|B_7^2|}b_B,\sqrt{|B_7^2|}b_K)
F_7(t_{7a})\right.
\nonumber\\
& &\left.+[1+r+(1-2r)x_K]H^{(b)}_{7}(Ab_B,C_7b_B,C_7b_K)F_7(t_{7b})\right]\;,
\end{eqnarray}
with
\begin{eqnarray}
& &B_7^2=(x_B-r^2)M_B^2\;,\;\;\;\; C_7^2=x_KM_B^2
\nonumber\\
& &t_{7a}=\max(A,\sqrt{|B_7^2|},1/b_B,1/b_K)\;,
\nonumber\\
& &t_{7b}=\max(A,C_7,1/b_B,1/b_K)\;,
\end{eqnarray}
where the function $F_7$ denotes the product
\begin{eqnarray}
F_7(t)=\alpha_s(t)c_7(t)\exp[-S(x_B,x_K,t,b_B,b_K)]\;.
\end{eqnarray}
The hard functions
\begin{eqnarray}
& &H^{(a)}_{7}(Ab_K,\sqrt{|B_7^2|}b_B,\sqrt{|B_7^2|}b_K)=
\nonumber\\
& &\hspace{1.0cm} K_0(Ab_K)
h(\sqrt{|B_7^2|}b_B,\sqrt{|B_7^2|}b_K)\;,\;\;\;\;B_7^2>0\;,
\nonumber\\
& &\hspace{1.0cm} K_0(Ab_K)
h'(\sqrt{|B_7^2|}b_B,\sqrt{|B_7^2|}b_K)\;,\;\;\;\;B_7^2<0\;,
\end{eqnarray}
with
\begin{eqnarray}
h&=&\theta(b_B-b_K)K_0(\sqrt{|B_7^2|}b_B)
I_0(\sqrt{|B_7^2|}b_K)+(b_B\leftrightarrow b_K)\;,
\nonumber\\
h'&=&i\frac{\pi}{2}\left[\theta(b_B-b_K)
H_0^{(1)}(\sqrt{|B_7^2|}b_B)J_0(\sqrt{|B_7^2|}b_K)+
(b_B\leftrightarrow b_K)\right]\;,
\end{eqnarray}
and 
\begin{eqnarray}
H^{(b)}_{7}(Ab_B,C_7b_B,C_7b_K)=K_0(Ab_B)
h(C_7b_B,C_7b_K)\;,
\end{eqnarray}
are derived from Figs.~2(a) and 2(b), respectively. The relation
$M_7^S=-M_7^P$ reflects the equality of the parity-conserving and
parity-violating contributions induced by $O_7$.

Four diagrams, Figs.~3(a)-(d), are associated with the operator $O_8$, where
the photon is radiated by each quark in the $B$ or $K^*$ mesons. The
corresponding amplitudes are
\begin{eqnarray}
M^S_8&=&-\Gamma^{(0)}\frac{1}{3}\int_0^1dx_Bdx_K\int_0^{1/\Lambda}
b_Bdb_Bb_Kdb_K\phi_B(x_B)\phi_{K^*}(x_K)
\nonumber\\
& &\times \Bigg\{(1-r^2+x_B)(rx_K+x_B)
H^{(a)}_8(Ab_K,B_8b_B,B_8b_K)F_8(t_{8a})
\nonumber\\
& &\hspace{0.5cm}+[(2-3r)x_K-x_B+r(1-x_K)(rx_K-2rx_B+3x_B)]
\nonumber\\
& &\hspace{1.0cm}\times
H^{(b)}_8(Ab_B,C_8b_B,C_8b_K)F_8(t_{8b})
\nonumber\\
& &\hspace{0.5cm}+(1+r)(1-r^2)[(1+r)x_B-rx_K]
\nonumber\\
& &\hspace{1.0cm}\times
H^{(c)}_8(\sqrt{|A'^2|}b_K,D_8b_B,D_8b_K)F_8(t_{8c})
\nonumber\\
& &\hspace{0.5cm}
-[(1-r^2)((1-r^2)(2-x_K)+(1+3r)(2x_K-x_B))
\nonumber\\
& &\hspace{1.0cm}
+2r^2x_K(x_K-x_B)]
\nonumber\\
& &\hspace{1.0cm}\times
H^{(d)}_8(\sqrt{|A'^2|}b_B,E_8b_B,E_8b_K)F_8(t_{8d})\Bigg\}\;,
\\
M^P_8&=&\Gamma^{(0)}\frac{1}{3}\int_0^1dx_Bdx_K\int_0^{1/\Lambda}
b_Bdb_Bb_Kdb_K\phi_B(x_B)\phi_{K^*}(x_K)
\nonumber\\
& &\times \Bigg\{(1-r^2+x_B)(rx_K+x_B)
H^{(a)}_8(Ab_K,B_8b_B,B_8b_K)F_8(t_{8a})
\nonumber\\
& &\hspace{0.5cm}+[(2-3r)x_K-x_B-r(1-x_K)(rx_K-2rx_B+3x_B)]
\nonumber\\
& &\hspace{1.0cm}\times
H^{(b)}_8(Ab_B,C_8b_B,C_8b_K)F_8(t_{8b})
\nonumber\\
& &\hspace{0.5cm}+(1-r)(1-r^2)[(1+r)x_B-rx_K]
\nonumber\\
& &\hspace{1.0cm}\times
H^{(c)}_8(\sqrt{|A'^2|}b_K,D_8b_B,D_8b_K)F(t_{8c})
\nonumber\\
& &\hspace{0.5cm}-[(1-r^2)((1-r^2)(2+x_K)-(1-3r)x_B)
\nonumber\\
& &\hspace{1.0cm}
-2r^2x_K(x_K-x_B)]
\nonumber\\
& &\hspace{1.0cm}\times
H^{(d)}_8(\sqrt{|A'^2|}b_B,E_8b_B,E_8b_K)F(t_{8d})\Bigg\}\;,
\end{eqnarray}
with
\begin{eqnarray}
& &A^{\prime 2}=(1-r^2)(x_B-x_K)M_B^2\;,\;\;\;\;
B_8^2=(1-r^2+x_B)M_B^2\;,
\nonumber\\
& &C_8^2=(1-x_K)M_B^2\;,\;\;\;\;
D_8^2=(1-r^2)x_BM_B^2\;,\;\;\;\; E_8^2=(1-r^2)x_KM_B^2\;,
\nonumber\\
& &t_{8a}=\max(A,B_8,1/b_B,1/b_K)\;,
\nonumber\\
& &t_{8b}=\max(A,C_8,1/b_B,1/b_K)\;,
\nonumber\\
& &t_{8c}=\max(\sqrt{|A'^2|},D_8,1/b_B,1/b_K)\;,
\nonumber\\
& &t_{8d}=\max(\sqrt{|A'^2|},E_8,1/b_B,1/b_K)\;.
\end{eqnarray}
and the funciton $F_8$,
\begin{eqnarray}
F_8(t)=\alpha_s(t)c_8(t)\exp[-S(x_B,x_K,t,b_B,b_K)]\;.
\end{eqnarray}
The hard functions
\begin{eqnarray}
& &H^{(a)}_{8}(Ab_K,B_8b_B,B_8b_K)=
H^{(b)}_{7}(Ab_K,B_8b_B,B_8b_K)\;,
\nonumber\\
& &H^{(b)}_8(Ab_B,C_8b_B,C_8b_K)=K_0(Ab_B)h'(C_8b_B,C_8b_K)
\nonumber\\
& &H^{(c)}_8(\sqrt{|A'^2|}b_K,D_8b_B,D_8b_K)=
\nonumber\\
& &\hspace{1.0cm} K_0(\sqrt{|A'^2|}b_K)h(D_8b_B,D_8b_K)\;,
\;\;\;\;A'^2\ge 0\;,
\nonumber\\
& &\hspace{1.0cm} i\frac{\pi}{2}H_0^{(1)}
(\sqrt{|A'^2|}b_K)h(D_8b_B,D_8b_K)\;,\;\;\;\;A'^2< 0\;,
\nonumber\\
& &H^{(d)}_8(\sqrt{|A'^2|}b_B,E_8b_B,E_8b_K)=
\nonumber\\
& &\hspace{1.0cm} K_0(\sqrt{|A'^2|}b_B)h'(E_8b_B,E_8b_K)\;,
\;\;\;\;A'^2\ge 0\;,
\nonumber\\
& &\hspace{1.0cm} i\frac{\pi}{2}H_0^{(1)}
(\sqrt{|A'^2|}b_B)h'(E_8b_B,E_8b_K)\;,\;\;\;\;A'^2< 0\;
\end{eqnarray}
are derived from Figs.~3(a)-(d), respectively.
It is obvious that the above factorization formulas bear the
features of Eq.~(\ref{for}).

The exponentials $\exp(-S)$ appearing in $M^i_l$
are the complete Sudakov form factors with the exponent
\begin{eqnarray}
S&=&s(x_BP_B^-,b_B)+2\int_{1/b_B}^{t}\frac{d\mu}{\mu}\gamma_q(\alpha_s(\mu))
\nonumber\\
& &+s(x_KP_K^+,b_K)+s((1-x_K)P_K^+,b_K)
+2\int_{1/b_K}^{t}\frac{d\mu}{\mu}\gamma_q(\alpha_s(\mu)).
\end{eqnarray}
The wave functions $\phi_B$ and $\phi_{K^*}$ satisfy the normalization,
\begin{equation}
\int_0^1\phi_i(x)dx=\frac{f_i}{2\sqrt{6}}\;,
\end{equation}
with the decay constant $f_i$, $i=B$ and $K^*$. The wave function for the
$K^*$ meson with transverse polarizations has been derived using QCD sum
rules \cite{BC}, which is given by
\begin{equation}
\phi_{K^*}=\frac{f_{K^*}}{\sqrt{6}}\frac{15}{4}(1-\xi^2)*[0.267(1-\xi^2)^2
+0.017+0.21\xi^3+0.07\xi]\;,
\end{equation}
with $\xi=1-2x$. As to the $B$ meson wave functions, we employ two models
\cite{YS,BW},
\begin{eqnarray}
\phi^{(I)}_B(x)&=&\frac{N_Bx(1-x)^2}{M_B^2+C_B(1-x)}\;,
\label{bw}\\
\phi^{(II)}_B(x)&=&N'_B\sqrt{x(1-x)}
\exp\left[-\frac{1}{2}\left(\frac{xM_B}{\omega}\right)^2\right]\;,
\label{os}
\end{eqnarray}
where $N_B$ and $N'_B$ are normalization constants; while 
$C_B$ and $\omega$ are shape parameters.

\section{RESULTS AND DISCUSSIONS}

In the evaluation of the various form factors and amplitudes, we adopt
$G_F=1.16639\times 10^{-5}$ GeV$^{-2}$, the flavor number $n_f=5$, the 
decay constants $f_B=200$ MeV
and $f_{K^*}=220$ MeV, the CKM matrix elements $|V^*_{ts}V_{tb}|=0.04$, the
masses $M_c=1.5$ GeV, $M_B=5.28$ GeV and $M_{K^*}=0.892$ GeV, the
${\bar B}^0$ meson lifetime $\tau_{B^0}=1.53$ ps, and the QCD scale
$\Lambda=0.2$ GeV \cite{CL}. We find that, no matter what value of the
shape parameter $C_B$ is chosen, the model $\phi^{(I)}_B$ in
Eq.~(\ref{bw}) with a flat profile leads to results smaller than
the CLEO data which gives ${\cal B}(B\to K^*\gamma)=
(4.2\pm 0.8 \pm 0.6) \times 10^{-5}$ \cite{CLEOEX2}. In fact,
the maximal prediction from $\phi^{(I)}_B$, corrsponding to the shape
parameter $C_B=-M_B^2$, is about $3.0\times 10^{-5}$, close
to the lower bound of the data.

On the other hand, using model $\phi^{(II)}_B$ in Eq.~(\ref{os}) with a
sharp peak at small $x$, we obtain a prediction much closer to the
experimental data. It is indeed found that, as $\omega=0.795$ GeV, a
prediction $4.204\times 10^{-5}$ for the branching ratio is reached,
which is equal to the central value of the experimental data.  
If varying the shape
parameter to both $\omega=0.79$ GeV and $\omega=0.80$ GeV, we obtain the
branching ratios $4.25\times 10^{-5}$ and $4.14\times 10^{-5}$,
respectively. It indicates that the allowed range for $\omega$ is wide due
to the yet large uncertainties of the data.

The detailed contribution from each amplitude $M^i_l$ is listed in Table
II in the unit of $10^{-6}$ GeV$^{-2}$. It is clear that $M^{S}_7$ and 
$M^P_7$ together 
give dominant contributions to the decay width.
One also sees that $M_2^S$ and $M_2^P$ are smaller by an order of magnitude, 
while the amplitudes associated with $O_8$ are highly suppressed. 
In table II, it is interesting to note that 
$M_2^S$ adds constructively to $M_7^S$, {\it i.e.},
$\vert M_2^S+M_7^S \vert ^2 > \vert M_7^S\vert ^2$. On the contrary, 
$M_2^P$ is destructive to $M_7^P$. Due to this cancellating effect, the 
inclusion of $O_2$ contribution only enhances the total rate by $2\%$
(This result is basically consistent with the estimation obtained in
\cite{MIL}). This result
is not sensitive to the choice of $\omega$. 
We obtain the same enhancement for the total rate
with $\omega$ chosen to be $0.79$ GeV
and $0.80$ GeV, respectively. 

By fitting our prediction for the branching ratio 
${\cal B}(B\to K^*\gamma)$ to
the CLEO data, we determine the $B$ meson wave function,
\begin{equation}
\phi_B(x)=0.740079\sqrt{x(1-x)}
\exp\left[-\frac{1}{2}\left(\frac{xM_B}{0.795\;\;{\rm GeV}}\right)^2
\right]\;,
\label{bwf}
\end{equation}
which possesses a sharp peak at the low momentum fraction $x$.
We stress that, however, Eq.~(\ref{bwf}) is not conclusive because of
the large allowed range of the shape parameter $\omega$. A more precise
$B$ meson wave function can be obtained by considering a global fit to the
data of various decay modes, including $B\to D^{(*)}\pi(\rho)$ \cite{LM}.
Once the $B$ meson wave fucntion is fixed, we shall employ it
in the evaluation of the nonleptonic charmless decays.

Finally, as mentioned in the Introduction, the authors of Ref.~\cite{GSW}
found that the diagrams without hard-gluon exchanges dominate over
Figs.~2(a) and 2(b) we have evaluated (only the operator $O_7$ was
considered in the calculation of the branching ratio 
${\cal B}(B\to K^*\gamma)$ in
\cite{GSW}): the latter contribute at most 23\% to the branching ratio, or
12\% to the decay amplitude. However, in that analysis, $\alpha_s$ is set
to 0.2, which is even smaller than $\alpha_s(M_B)=0.23$ evaluated at the
$B$ meson mass. We argue that such a small coupling constant is
inappropriate, since the momentum flow involved in the decay process is most
likely less than the $b$ quark mass $M_b$, say, roughly $1\sim 2$ GeV, which
corresponds to $\alpha_s\approx 0.4$. The $K^*$ meson mass was neglected in
\cite{GSW}, such that Fig.~2(a) does not contribute. In our analysis we did
not make this approximation, and observed that the contribution from
Fig.~2(a) is about $M_{K^*}/M_B\approx 1/5$ of that from Fig.~2(b).
Moreover, a flat $B$ meson wave function corresponding to the shape
parameter $\omega\approx 1.3$ GeV was adopted in the computation of the
higher-order contributions in \cite{GSW}, which is far beyond
$\omega \approx 0.4$ GeV specified in \cite{BW}. If a sharper $B$ meson wave
function is employed, these contributions will be enhanced at least by a
factor of 3. Note that the leading-order contributions considered in
\cite{GSW} are less sensitive to the variation of $\omega$ in the wave
function. Adding up the above enhancements, the amplitude for
$O(\alpha_s)$-corrections becomes approximately equal to
the leading-order contribution. In this sense the analysis
in \cite{GSW} does not seem to be self-consistent. 

In our approach, the
momentum of the spectator quark in the $B$ meson is parametrized
into the minus direction, while the momentum of the spectator quark in
the $K^*$ meson is parametrized in the plus direction. This
parametrization is appropriate because of the hard gluon exchange. Note
that the factorization formulas presented in Sect. III are constructed
based on the diagrams with at least one hard gluon exchange. For
example, the infrared divergences from self-energy corrections to
the spectator quark are factorized into the $B$ meson wave function,
if they occur before the hard gluon exchange; and into the $K^*$ meson
wave function, if they occur after the hard gluon exchange. Without hard
gluons to distinguish the initial and final states, factorization of
self-energy corrections to the spectator quark is ambiguous. Therefore,
the diagrams without hard gluons do not appear in the regime of PQCD
factorization theorems, and those with one hard gluon are indeed leading.
If ignoring the validity of the factorization, we may proceed with
evaluating the contribution of $O_7$ from the diagram without hard
gluons in the PQCD framework. It is trivial to obtain the hard
subamplitude in momentum space,
\begin{equation}
H_7^{(0)}=\frac{(1+r)(1-r^2)}{2M_B}\delta^3(k_1-k_2)\;,
\end{equation}
where the $\delta$-function requires that the longitudinal momenta
of the spectrator quarks in the $B$ and $K^*$ mesons are in the same
direction. Fourier transforming the above expression into $b$ space, and 
convoluting it with the wave funcitons in Eqs.~(35) and (\ref{bwf}) and
with the Sudakov factor, we derive the amplitudes
\begin{eqnarray}
M_7^{S(0)}=-M_7^{P(0)}&=&\Gamma^{(0)}\int_0^1
dx\int_0^{1/\Lambda}bdb\phi_B(x)\phi_{K^*}(x)
\nonumber\\
& &\times \frac{(1+r)(1-r^2)}{4\pi^2 M_B^2 C_F}c_7(1/b)
\exp[-S(x,x,1/b,b,b)]\;.
\label{m70}
\end{eqnarray}
Without hard gluons, the momentum fraction $x$ and the transverse
extent $b$ are equal for the $B$ and $K^*$ mesons. We have set the
hard scale $t$ to $1/b$ due to the lack of gluon momentum transfer.
A simple numerical work on Eq.~(\ref{m70}) gives
$M_7^{S(0)}/\Gamma^{(0)}=-2.51\times 10^{-6}$ GeV$^{-2}$, which is of 
the same order as the contributions from $O_2$ and $O_8$. One of
the reasons for the smallness of $M_7^{S(0)}$, compared to the values
obtained in \cite{GSW}, is the additional strong Sudakov suppression.
Referred to $M_7^S$ listed in Table II, $M_7^{S(0)}$ is negligible.
If including $M_7^{S(0)}$ and $M_7^{P(0)}$, the branching ratio 
$B(B\to K^*\gamma)$ will increase by only 1.7\%.

\section{CONCLUSION}

In this paper we have extended the PQCD three-scale factorization theorem
to the penguin-induced radiative decay $B\to K^*\gamma$, which takes into
account the Sudakov resummation for large logarithmic corrections to
this process. We have included the non-spectator contribution from the 
current-current operator $O_2$ besides the standard contribution 
given by magnetic-penguin operators $O_7$ and $O_8$. It turns out that the 
contribution by $O_2$ is negligible due to certain cancellations.  
The contributions 
from $O_8$ and other operators in the 
effective Hamiltonian are also quite small.
Finally, we have determined the $B$ meson wave function from the best fit to
the experimental data of ${\cal B}(B\to K^*\gamma)$, which will be employed 
to make predictions of other $B$ meson decays.

\acknowledgments
We thank X.-G. He and W.-S. Hou for discussions.
This work was supported in part by the National Science Council of R.O.C.
under the Grant Nos. NSC-88-2112-M-006-013 and NSC-88-2112-M-009-002.

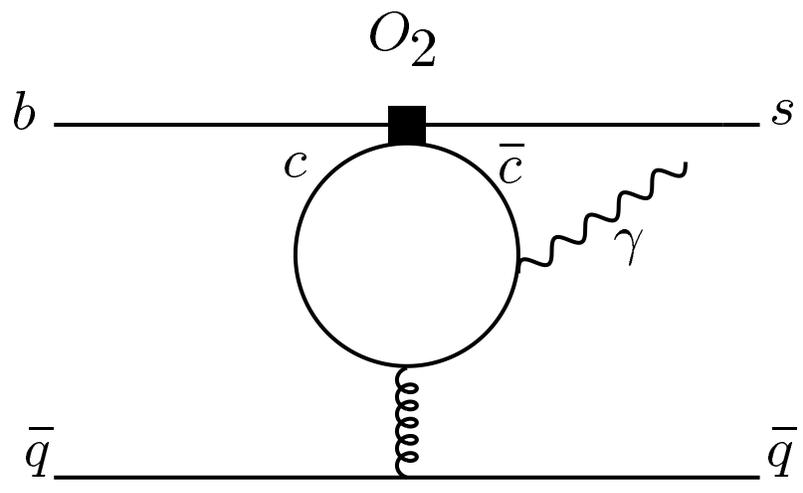
\begin{figure}
\caption{Contributions to the $B\to K^*\gamma$ decay from
the current-current operator $O_2$. The diagram with a
photon emitted from the other side of the charm-quark loop is not shown.}
\label{fig1}

\end{figure}    

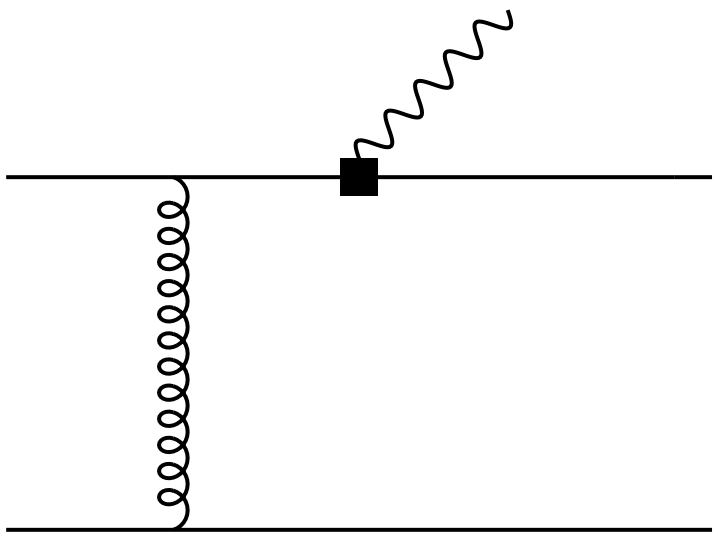
\begin{figure}
\caption{Contributions to the $B\to K^*\gamma$ decay from the 
magnetic-penguin operator $O_7$.}
\label{fig2}

\end{figure}  

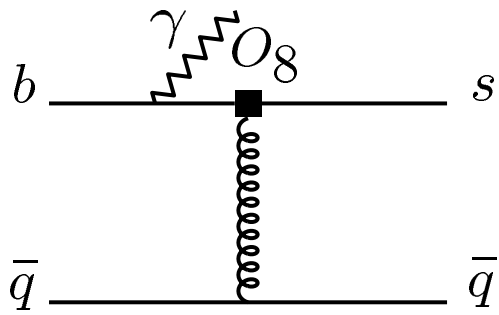
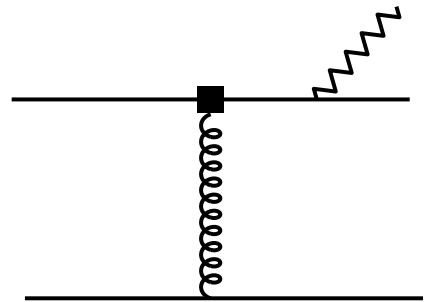
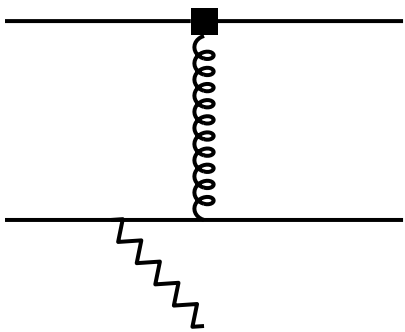
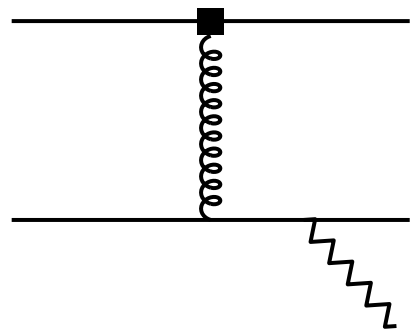
\begin{figure}
\caption{Contributions to the $B\to K^*\gamma$ decay from the
chromo-magnetic-penguin operator $O_8$.}
\label{fig3}

\end{figure}  

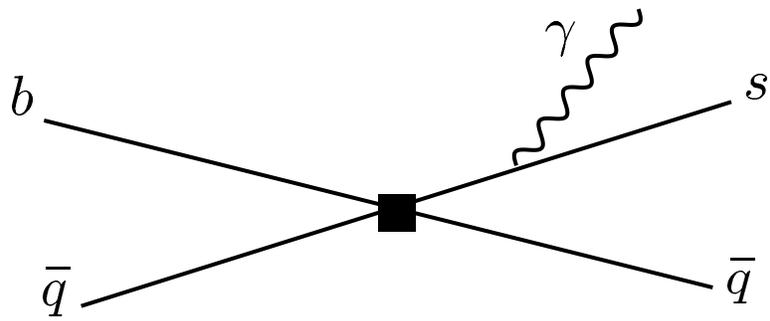
\begin{figure}
\caption{Contributions to the $B\to K^*\gamma$ decay from the 
strong-penguin operators $O_3, O_4, \cdots O_6$. The dark square denotes 
insertions of the operators $O_3, O_4, \cdots O_6$.}
\label{fig4}

\end{figure}  
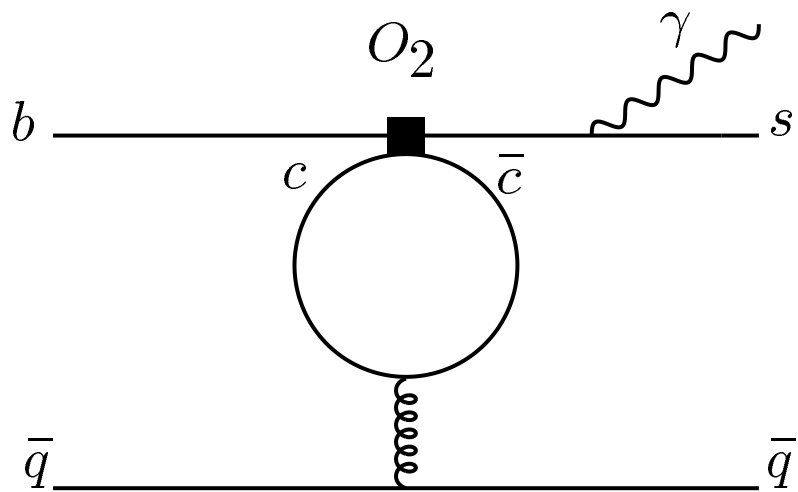
\begin{figure}
\caption{Contributions to the $B\to K^*\gamma$ decay from an 
$O_2$-insertion and a {\it bremsstrahlung} photon.}
\label{fig5}

\end{figure}  

\newpage
Table I. Hard subamplitudes obtained from Figs.~1-3. The quantities
${\tilde A}_4$ and ${\tilde A}_5$ are integrands of $I_{11}$ and
$I_{20}-I_{10}$ respectively, where the general integral
$I_{ab}$ is defined in Eq. (10).
\vskip 1.0cm

\[ \begin{array}{rc}   \hline\hline\\
{\rm Diagram} & H^S \\
\hline   \\
O_2     &{\displaystyle
\frac{4}{3}\frac{(1-r)(1-r^2)x_K[(1-r^2+2rx_K+2x_B){\tilde A}_4+
(rx_K+3x_B){\tilde A}_5]}{x_Kx_Bm_B^2+({\bf k}_{KT}-{\bf k}_{BT})^2}} \\
        &    \\
O_7(a)     &{\displaystyle
\frac{2r(1-r^2)}{[x_Kx_Bm_B^2+({\bf k}_{KT}-{\bf k}_{BT})^2]
[(x_B-r)m_B^2+k_{BT}^2]} } \\
        &    \\
O_7(b)   &  {\displaystyle
\frac{2(1-r^2)[1+r+(1-2r)x_K]}{[x_Kx_Bm_B^2+({\bf k}_{KT}-{\bf k}_{BT})^2]
(x_Km_B^2+k_{KT}^2)} } \\
        &   \\
O_8(a)   &  {\displaystyle
-\frac{(1-r^2+x_B)(rx_K+x_B)}
{3[x_Kx_Bm_B^2+({\bf k}_{KT}-{\bf k}_{BT})^2]
[(1-r^2+x_B)m_B^2+k_{BT}^2]} } \\
        &   \\
O_8(b)   &  {\displaystyle
-\frac{(2-3r)x_K-x_B+r(1-x_K)(rx_K-2rx_B+3x_B)}
{3[x_Kx_Bm_B^2+({\bf k}_{KT}-{\bf k}_{BT})^2]
[(x_K-1)m_B^2+k_{KT}^2]} } \\
        &    \\
O_8(c)   &  {\displaystyle
-\frac{(1+r)(1-r^2)[(1+r)x_B-rx_K]}
{3[(1-r^2)(x_B-x_K)m_B^2+({\bf k}_{KT}-{\bf k}_{BT})^2]
[(1-r^2)x_Bm_B^2+k_{BT}^2]} } \\
        &    \\
O_8(d)   & {\displaystyle
\frac{(1-r^2)[(1-r^2)(2-x_K)+(1+3r)(2x_K-x_B)]+2r^2x_K(x_K-x_B)}
{3[(1-r^2)(x_B-x_K)M_B^2+({\bf k}_{KT}-{\bf k}_{BT})^2]
[(r^2-1)x_Km_B^2+k_{KT}^2]} } \\
   & \\
\hline
\end{array}   \]

\[ \begin{array}{rc}   \hline\hline\\
{\rm Diagram} & H^P \\
\hline   \\
O_2     &{\displaystyle
\frac{4}{3}\frac{(1-r^2)x_K[(1-r)(1-r^2)+2r^2x_K+2x_B){\tilde A}_4+
(r(1+r)x_K+(3-rx_B)){\tilde A}_5]}
{x_Kx_Bm_B^2+({\bf k}_{KT}-{\bf k}_{BT})^2}} \\
        &    \\
O_7(a)     &{\displaystyle
-\frac{2r(1-r^2)}{[x_Kx_Bm_B^2+({\bf k}_{KT}-{\bf k}_{BT})^2]
[(x_B-r)m_B^2+k_{BT}^2]} } \\
        &    \\
O_7(b)   &  {\displaystyle
-\frac{2(1-r^2)[1+r+(1-2r)x_K]}{[x_Kx_Bm_B^2+({\bf k}_{KT}-{\bf k}_{BT})^2]
(x_Km_B^2+k_{KT}^2)} } \\
        &   \\
O_8(a)   &  {\displaystyle
\frac{(1-r^2+x_B)(rx_K+x_B)}
{3[x_Kx_Bm_B^2+({\bf k}_{KT}-{\bf k}_{BT})^2]
[(1-r^2+x_B)m_B^2+k_{BT}^2]} } \\
        &   \\
O_8(b)   &  {\displaystyle
\frac{(2-3r)x_K-x_B-r(1-x_K)(rx_K-2rx_B+3x_B)}
{3[x_Kx_Bm_B^2+({\bf k}_{KT}-{\bf k}_{BT})^2]
[(x_K-1)m_B^2+k_{KT}^2]} } \\
        &    \\
O_8(c)   &  {\displaystyle
\frac{(1-r)(1-r^2)[(1+r)x_B-rx_K]}
{3[(1-r^2)(x_B-x_K)m_B^2+({\bf k}_{KT}-{\bf k}_{BT})^2]
[(1-r^2)x_Bm_B^2+k_{BT}^2]} } \\
        &    \\
O_8(d)   &  {\displaystyle
-\frac{(1-r^2)[(1-r^2)(2+x_K)-(1-3r)x_B]-2r^2x_K(x_K-x_B)}
{3[(1-r^2)(x_B-x_K)M_B^2+({\bf k}_{KT}-{\bf k}_{BT})^2]
[(r^2-1)x_Km_B^2+k_{KT}^2]} } \\
  & \\
\hline
\end{array}   \]

\newpage
Table II. The amplitudes $M^i_l$ in the unit of $10^{-6}$ GeV$^{-2}$.
\vskip 1.0cm

\[\begin{array}{ccc}\hline
M^S_2/\Gamma_0 & M^S_7/\Gamma_0 & M^S_8/\Gamma_0 \\ 
\hline
-2.46-14.17i & -140.14-143.31i & -0.58+1.10i \\
\hline
M^P_2/\Gamma_0 & M^P_7/\Gamma_0 & M^P_8/\Gamma_0 \\ 
\hline
-1.21-11.05i & 140.14+143.31i & 0.54-1.33i \\
\hline 
\end{array}\]


\begin{thebibliography}{99}
\bibitem{SIGN}
The signs of $O_7$ and $O_8$ are consistent with the 
covariant derivative $D_{\mu}=\partial_{\mu}+igT^aA_{\mu}^a+ieQA_{\mu}$,
where $A_{\mu}^a$ and $A_{\mu}$ denote gluon and photon
fields respectively.
\bibitem{ORDER}
Strictly speaking, the order of this diagram is $eG_F(A_1\alpha_s
\ln{M_W\over \mu}+B_1\alpha_s+A_2\alpha_s^2
\ln^2{M_W\over \mu}+B_2\alpha_s^2
\ln{M_W\over \mu}+\cdots)$, where coefficients 
$A_i$ and $B_i$ represent contributions from leading and next-to-leading
logarithms.  We simply count the order according to 
the first two terms in the series, $A_1\alpha_s
\ln{M_W\over \mu}$ and $B_1\alpha_s$.
Same rule of counting applies to diagrams in Fig.~1-3.
\end{thebibliography}
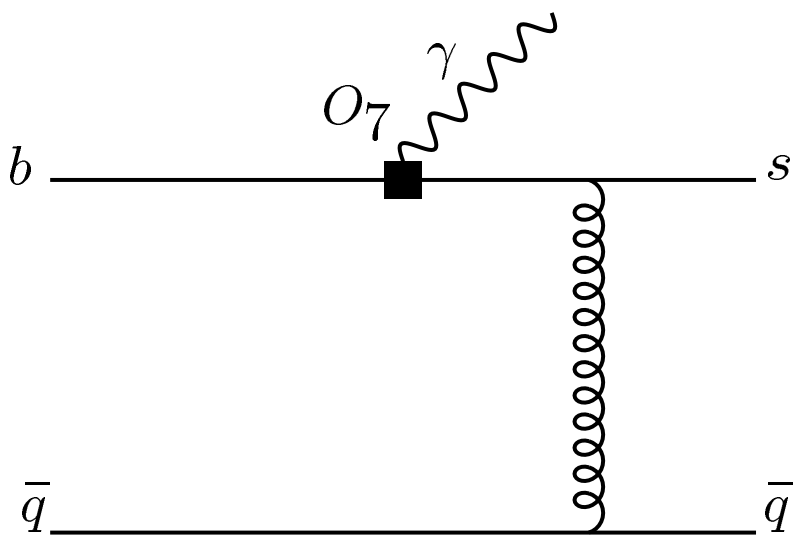
\end{document}